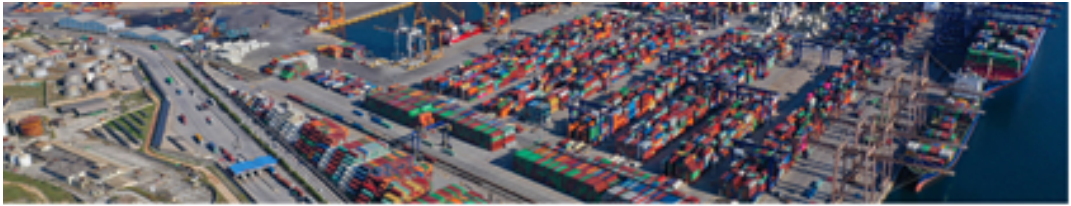

# Framework Artifact for the Road-Based Physical Internet based on Internet Protocols

Steffen Kaup[1], André Ludwig[2], Bogdan Franczyk[1,3]

1. Leipzig University, Information Systems Institute, Leipzig, Germany
2. Kühne Logistics University, Computer Science in Logistics, Hamburg, Germany
3. Wrocław University of Economics, Wrocław, Poland

*{kaup\*, franczyk}@wifa.uni-leipzig.de, andre.ludwig@the-klu.org*

*Abstract*

The Physical Internet (PI, π) raises high expectations for efficiency gains in transport and logistics. The PI represents the network of logistics networks for physical objects in analogy to the Data Internet (DI). Road based traffic represents one of these logistics networks. Here, many empty runs and underutilized trips still take place (International Transport Forum, 2019). Hence, there is a lot of potential in the road-based Physical Internet (RBPI), which will have an impact on transport and logistics strategies, but also on vehicle design. On the DI, logistics strategies are implemented in protocols. In order to transfer such concepts to the RBPI, relevant protocols of the DI had been analyzed and transferred to the world of physical objects. However, not all functionalities can be transferred one-to-one, e.g. a data packet in the DI can simply be re-generated by a hub in case of damage or loss. To compensate for the challenges, a framework artifact has been designed with appropriate transformation customizations based on design science principles (vom Brocke, 2007). From this, resulting requirements for future vehicles were derived. This paper makes a contribution to the implementation of the RBPI in order to fit road based vehicles to the future world of transport and logistics.

## 1. Introduction

The Physical Internet (PI) represents the network of logistics networks for physical objects in analogy to the Data Internet (DI). 72 percent of freight traffic in Germany is done by trucks on the road (Statistisches Bundesamt, 2019). This is why the road-based Physical Internet (RBPI) plays a central role within the PI. The capacity for the movement of goods on the road is restricted by physical constraints, such as the number and available volume of vehicles, as a part of the π-movers family, and limited road connections. Here, many empty runs and underutilized trips still take place (International Transport Forum, 2019). Through a road-based Physical Internet (RBPI), methods of a very established information network regarding resilience and efficiency, the Data Internet (DI), are transferred to the world of freight road transport. As a vision of the future, freight is being navigated through an existing transport network in a way analogous to how information packets are transferred from a host computer to another computer on the DI, as visualized in Figure 1.

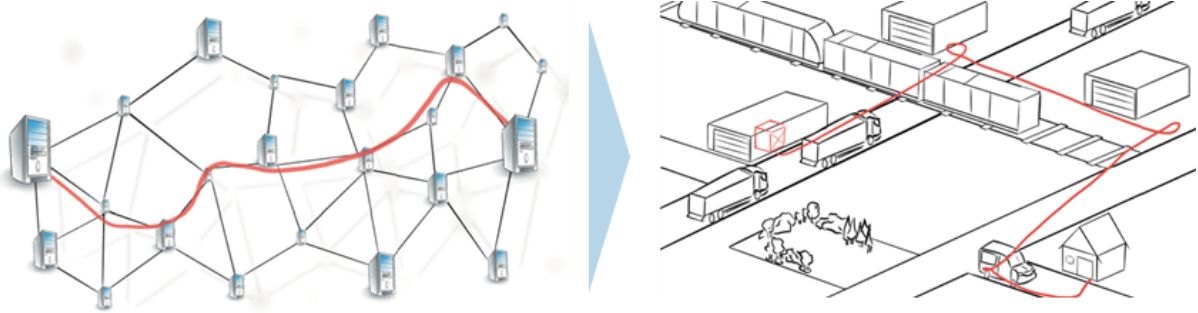

*Figure 1: Transformation from DI to RBPI (own visualization)*

On the DI, data transport methods are covered by Internet Protocols (Badach and Hoffmann, 2007). For the RBPI, an adequate concept is still missing. However, this concept as well, like the basic idea of the DI itself, can be adapted from the existing Internet Protocol family, as transferred in the following sections. After presenting related work on the topic in Section 2, research methodology is introduced in Section 3. Subsequently, Section 4 starts with the transformation of Internet Protocols and proceeds to develop a design artifact for the RBPI. In Section 5, conclusions are drawn and further steps on the road to implementation of the RBPI are suggested.

## 2. Related Work

Research related to the Physical Internet dates back to the year 2009, initiated by the idea of Benoit Montreuil to transform the Digital Internet to a Physical Internet. He was inspired by an article which bore this headline (Markillie, 2006). Together with Ballot and Meller he wrote a manifesto that describes a lot of facets of this transformation (Montreuil, Meller and Ballot, 2010). Since this initial publication, research has developed further through numerous contributions at conferences, such as the 'International Physical Internet Conference' (IPIC). Hardly any papers published there dealt with the research of protocols and if so, only on an abstract level (Zikria *et al.*, 2018). Fundamental work concerning network-based software architecture was carried out in a dissertation (Fielding, 2000). A conceptual approach to solve the problem of finding a feasible route with the lowest total cost and an appropriate time is designed with a graph based model (Dong and Franklin, 2020). A model transfer was carried out within the context of crowd logistics with regard to available cargo space capacities of current road traffic data (Kaup and Demircioglu, 2017). In recent years, two industrial standards for the Internet of Things (IoT) have developed, the 'Reference Architecture Model Industrie 4.0' (RAMI 4.0) and the 'Industrial Internet Reference Architecture' (IIRA). The models aim to capture production objects throughout the entire life cycle and to map them uniformly and consistently on the IT side. The differences between RAMI 4.0 and IIRA lie in the different emphasis in scope and depth. The layer of the RAMI model responsible for organising the IoT components is based on the Open Systems Interconnection (OSI) model (Shi-Wan *et al.*, 2017). Hence, our research focuses on the OSI model. Furthermore, a recent study recommends the detailed analysis of Internet Protocols in order to sharpen the analogy from DI to PI (van Luik *et al.*, 2020).

*2.1 Open Systems Interconnection (OSI) model and existing transformations*

IT network services have been structured into a seven layer service model, the Open Systems Interconnection (OSI) model. OSI is a reference model for network protocols as a layered architecture. It has been published as a standard by the International Organization for Standardization (ISO) since 1984. For each layer, functions and protocols are defined which have to fulfill certain tasks within the communication between two systems (Kaup and Neumayer, 2003). The layers 1-3 represent network-oriented functions, like transmission and switching. Layer 4 is called the transport layer and is intended to enable the different transport networks to be used for the connection of end-to-end systems. The protocols of layers 5-7 are application-specific. Montreuil et al. (Montreuil, 2012) introduced an Open Logistics Interconnection model (OLI) as an analogy to its digital equivalent OSI. The OLI model proposes the seven layers: physical (1), link (2), network (3), routing (4), shipping (5), encapsulation (6) and logistics web (7), as shown in Figure 2.

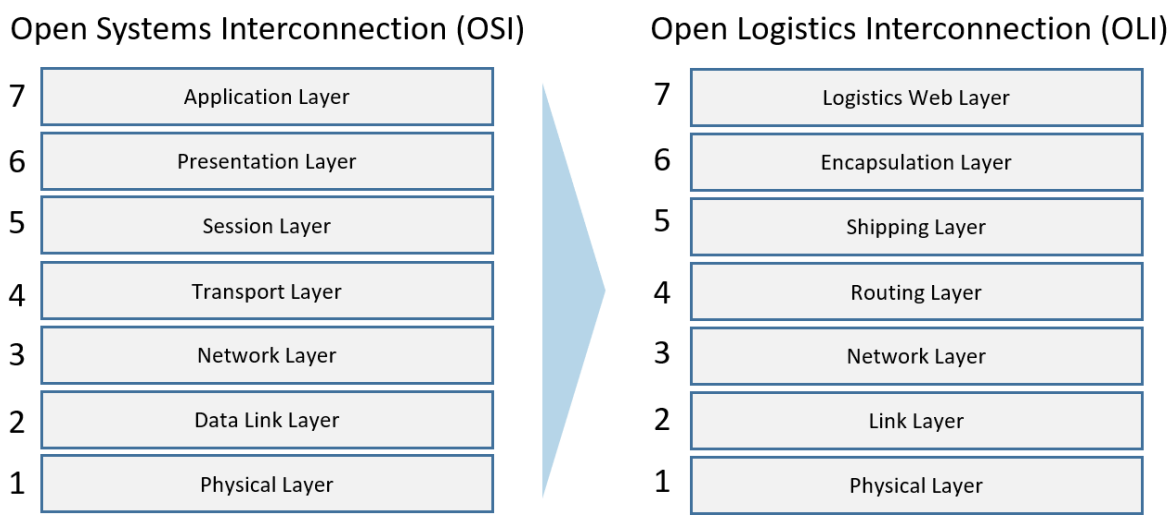

*Figure 2: Open Logistics Interconnection model, adapted from (Montreuil, Meller and Ballot, 2010)*

A working group of the European research group 'Alliance for Logistics Innovations through Collaboration in Europe' (ALICE) proposed to concentrate on a conceptual five layers model (Liesa *et al.*, 2020), shown in Table 1, which goes along with the latest research (Dong and Franklin, 2020).

*Table 1: Physical Internet Conceptual Five Layers Model (Liesa et al., 2020)*

| Protocol Layer | Description |
|---|---|
| "Application" Layer | This layer is where the actual goods are "defined" and human readable information about the goods is created. It is in this layer that this information and these goods are prepared for transmission/transport to their destination. As with the Internet, this packet of information and the associated physical goods (the shipment) is our "message." |
| "Transport" Layer | At the transport layer shipments are broken up into sizes that are transportable by standard sized containers or network defined standard transport mechanisms (for goods not amenable to standard containers). In addition, the transport layer provides services that ensure delivery of the shipment and manage flows between the sending location and destination. The standard loads that are shipped out from the transport layer are our "segments." |
| "Network" Layer | The network layer takes the "segments" constructed in the transport layer and manages all services required to deliver these "segments" to their destination. This layer defines how all nodes between source and destination should respond to handling and controlling the goods that are in the segments. The information concerning handling and control of the segments is attached to the segment and the combination of this information and the shipment segment forms our "datagram." |
| "Link" Layer | The link layer takes the "datagram" from the network layer and passes it from the current node to the next node in the network. The services that the link layer provides depends on the mode of transport between nodes. The encapsulated "datagram," which includes all information on how the particular transport mode is to handle the shipment, is called a "frame." |
| "Physical" Layer | The physical layer of the Physical Internet actually moves the "bits" of a shipment between the linked nodes. The services provided are both link and mode dependent and depend heavily on mode, carrier, regulatory bodies, etc. |

Building upon this research, this paper conducts a deep dive into the five transport relevant layers, as developed by (Liesa *et al.*, 2020). From the results of the protocol analysis, functionalities and attributes are transformed to the RBPI, weaknesses of a one-to-one transfer are identified and possible fixes discussed and adopted as possible.

## 2.2 Knowledge gap and research questions

With the PI, methods of a very established information network regarding resilience and efficiency, the DI, are transferred to physical goods transport. Some concepts have been developed so far, but without sufficient depth in terms of protocols. These protocols implement the methods within the DI. This paper analyzes Internet Protocols in depth and dares the transformation of their functionalities and attributes to the RBPI. This transformation answers the leading Research Question (RQ): *'What are the corresponding functionalities and attributes within the road-based Physical Internet resulting from the analysis of Digital Internet protocols?'* that is solved with the following Sub-Questions (SQ):

- SQ$_1$: Which functions and/or attributes can be transferred one-to-one from the DI protocols to the RBPI?
- SQ$_2$: Which of the non-transferable functions and/or attributes require replacement?
- SQ$_3$: What vehicle requirements can be derived from the protocol transformation?

The following Section describes relevant principles to solve these design questions.

## 3. Research Methodology

The purpose of this paper is to transfer the operating principles of the DI to the world of road-based freight transport. Hence, this is a design problem, even though design science research (DSR) has a methodological answer: design principles for reference modelling. Basically, five design principles are distinguished: analogy, specialization, aggregation, instantiation and configuration (vom Brocke, 2007). Each principle represents a special technique of reusing methods or content from the original model in order to build a target model.

*Table 2: Principles for reuse in DSR reference modelling, adapted from (vom Brocke, 2007)*

| Principle | Visualization | Description |
|---|---|---|
| Configuration | 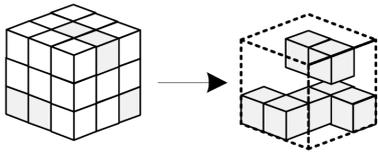 | The derived model is a selection of relevant properties and methods of the original model. |
| Instantiation | 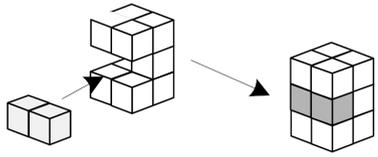 | Instantiation offers the opportunity to construct models for which both the methods as well as the properties are reusable. |
| Aggregation | 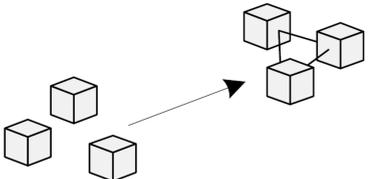 | Aggregation offers the potential of combining model statements of original models in new contexts. |
| Specialization | 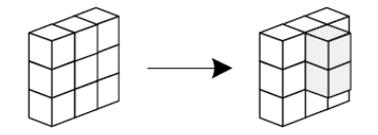 | Specialisation allows the taking over of general construction methods and/or properties and extending them to specific demands. |
| Analogy | 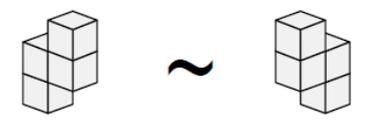 | The relation between the original model and the resulting model is based on a perceived similarity of both models regarding a certain aspect. |

The similarity of the two models, original and target model, depends on the selection and arrangement of components, methods and properties within the systems (Fielding, 2000). The principles move further and further away from the original concept from top to bottom, i.e. starting with 'configuration' and ending with 'analogy'. While 'configuration' merely involves the configuration of existing components, and many methods and properties are reused on a one-to-one basis in 'instantiation', the principle of 'analogy' only leads to a roughly corresponding solution. The principle of transformation has to be chosen in such a way that the transferred method from the DI works in the targeted world, the RBPI. In the following Section, properties and methods of the OSI layers are analyzed in respect of reuse regarding the RBPI.

**4. From DI to RBPI: Protocol Transformation**

This Section builds on the research of the sense project (Liesa *et al.*, 2020), which identified five relevant layers for transformation. In contrast to the OLI model (Montreuil, Meller and Ballot, 2010), layers 5-8 are combined into one layer. This layer then also contains the protocols responsible for routing mechanisms, as Figure 3 shows.

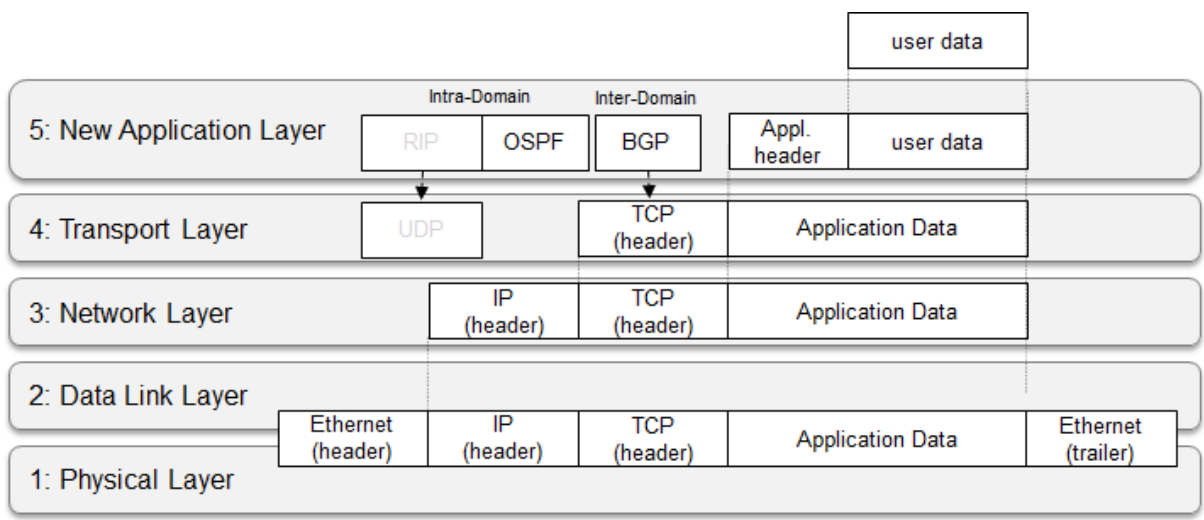

*Figure 3: Internet Protocols based on the five layer model (own visualization)*

In the following Subsections, the protocols and their functions and attributes are analyzed with respect to their reusability and transformation for the RBPI.

**Transformation of Layer 5: The aggregated Application Layer (OSI Layers 5-7)**

We start with the new aggregated application layer (Liesa *et al.*, 2020), where the transport journey begins. *'It is in this layer that this information and these goods are prepared for transmission/transport to their destination. (Liesa et al., 2020)'*. This information contains sender and destination addresses, treatment requirements, latest delivery date and the cost budget. This information aggregation in combination with the associated physical goods, or so-called shipment, is the counterpart of a message within the RBPI. On this layer, the protocols RIP (Routing Information Protocol), OSPF (Open Shortest Path Information First) and BGP (Border Gateway Protocol) implement different routing strategies, as visualized in Figure 6. The RIP protocol is a distance vector protocol. This means that decisions are made only on the basis of the number of hubs (or so-called π-nodes) on possible connection routes. The route with the lowest number of interconnected hubs wins, regardless of whether there is traffic with free capacity here at the desired time. It is based on the Used Datagram Protocol (UDP) and works connectionless, i.e. for each sub-component this routing process takes place again instead of maintaining a connection for a number of transmissions. There are approaches to empower UDP to become connection-oriented in order to be nearly equivalent to a TCP connection, e.g. via QUIC (Kumar, 2020). The advantages lie in reduced latency for multiplex connections, which are, however, negligible for the RBPI.

The OSPF and BGP protocols work differently. They provide the network nodes with routing tables containing current network information. With OSPF, there are so-called autonomous systems that synchronize with each other. BGP, on the other hand, provides an overarching exchange, which makes it particularly suitable for cross-system exchange of routing information via routing tables as shown in Figure 4.

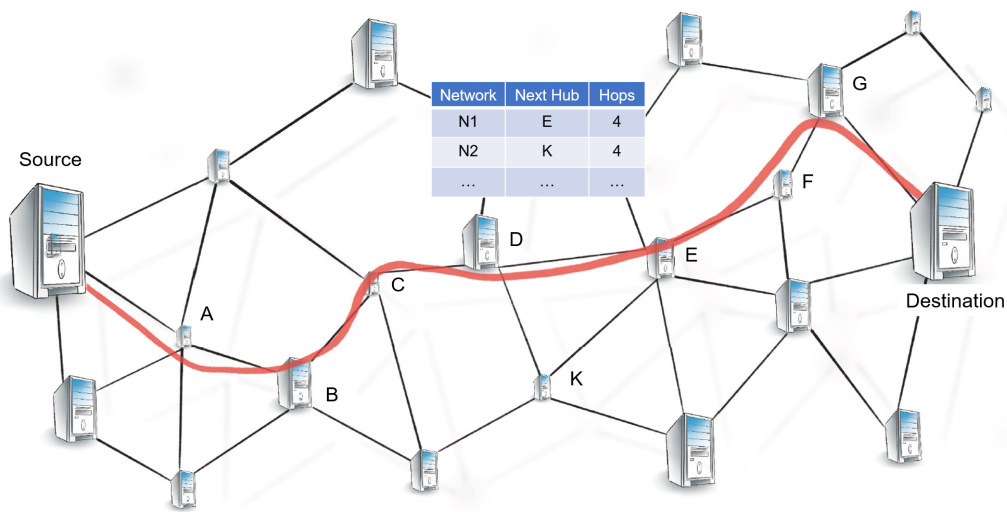

*Figure 4: Operation of the BGP-Protocol, based on (Kaup, Ludwig and Franczyk, 2020)*

The BGP-Protocol is based on the Transmission Control Protocol (TCP) on Layer 4 and the Internet Protocol (IP) on Layer 3. For this reason, only the TCP/IP protocol family was considered by the SENSE group (Liesa *et al.*, 2020), which is again confirmed by the evaluation of the routing protocols on Layer 5.

**Transformation of Layer 4: From Transport Layer to Routing Layer**

The Transport Layer ensures that all data is transmitted completely and arrives correctly at the receiver. '*At the transport layer shipments are broken up into sizes that are transportable by standard sized containers or network defined standard transport mechanisms*' *(Liesa et al., 2020)*. In addition, the transport layer provides services that ensure the faultless delivery of the shipment and manage the flows between the sending location and the corresponding destination. For this purpose, the segments are numbered sequentially to prevent double transmission of freight packets and to reconstruct their correct sequence at the destination side, as described and transferred in Table 3 with the principle of analogy.

*Table 3: Transport Layer Transformation Table*

| Task within DI | Principle | Task within RBPI |
| --- | --- | --- |
| **Segmentation and reassembly** | | |
| A message is divided into transmittable segments. Each segment contains a sequence number. These sequence numbers enable the transport layer to reassemble the message correctly upon arriving at the destination and to identify and replace packets that were lost in transmission. | Analogy | An order can be divided into individual sub-components, e.g. bulky or heavy parts and small or low-weight parts. These components can take different routes through the RBPI to their destination. If freight is decomposed into sub-components, these sub-components and their order must be provided with instructions for reassembly by the responsible π-node. |
| **Flow and error control** | | |
| This method ensures faultless operation, see also Table 3: Acknowledge and checksum attributes. | Analogy | It is not trivial to implement something like a checksum mechanism to automatically identify damaged packages. This can be done almost exclusively by visual inspection. |

The tasks described in Table 3 are implemented by the TCP-Protocol. Its methods and properties are transferred in detail as follows:

*Table 4: Transformation of TCP packet header to OLI TCP header*

| TCP attribute | # bits | Function within DI | Function within RBPI |
| --- | --- | --- | --- |
| Source/ Destination | 16/16 | Represents the port addresses of source and destination. | Ports could be used to identify a π-node's point of entry or exit. |
| Sequence number | 32 | The sequence number indicates the position of the data packet in the complete message. | The *sequence number* is required for the composition of the π-containers belonging to the same shipment. |
| Acknowledge-ment number | 32 | When the Acknowledge flag (ACK) is activated, the next sequence number is written in this field. | The *following sequence number* is of interest if some components are 3D-printed on the last π-node instead of being transmitted. |

| TCP attribute | # bits | Function within DI | Function within RBPI |
|---|---|---|---|
| Data offset | 4 | Indicates the size of optional information. | Indicates the size of all transport relevant information. |
| Reserved | 3 | For future use and should be set to zero. | Not relevant, can be used for additional transport information. |
| ECN/NS/CWR | 1/1/1 | The Congestion Experienced Flag (ECN) serves as an indication of network congestion. | This flag could be used to indicate identified (or impending) congestions on specific routes. |
| URG | 1 | Indicates the significance of the URGent pointer field. | Indicates an *urgent* or *very valuable* freight item. |
| ACK | 1 | Indicates significance of the acknowledgement number. | *Acknowledgement information of delivery* is requested. |
| PSH | 1 | Trigger for forwarding buffered data to the receiving application. | negligible |
| RST | 1 | Resets the connection. | negligible |
| SYN/FIN | 1/1 | Indicates the first (SYN) or last (FIN) packet within a transmission. | Indicates the first/last π-container of a shipment with more than one π-container. |
| Window size | 16 | The number of bytes that the sender of this segment is currently ready to receive. | Size of the storage location at the π-node of the receiver. |
| Checksum | 16 | The 16-bit checksum field is used for error checking the header, the payload and a pseudo-header. | Damaged packages can be identified almost exclusively by visual inspection. |
| Urgent pointer | 16 | Points to the last urgent data byte. | No one-to-one transfer is obvious. |
| Options | 32 | Optional information | Optional information |

**Transformation of Layer 3: The Network Layer**

The Network Layer describes procedures for exchanging files between addressable systems. The most important tasks of the network layer include the provision of cross-network addresses (IP), the routing or creation and updating of routing tables and the fragmentation of data packets (Badach, Hoffmann and Knauer, 1997). Hence, this layer *'defines how all nodes between source and destination should respond to handling and controlling freight contained in the segments' (Liesa et al., 2020)*, as detailed in the following table.

*Table 5: Network Layer Transformation Table*

| Task within DI | Principle | Task within RBPI |
|---|---|---|
| **Routing** | | |
| The router decides the next appropriate node based on the destination address and the use of routing tables. | Analogy | Determine the next π-node, depending on relevant information about the current traffic situation stored in the current π-node. Individual components of a delivery can take different routes to the destination. |
| **Logical Addressing** | | |
| If a packet passes the network boundary, another addressing system is necessary to help distinguish the source and destination systems. | Specialization | Different networks could be represented by different shipping companies with different address spaces. This would then be the task of the π-node to mediate. |

Each involved element, such as servers and terminals in the DI, has a unique IP address. The first major version of IP, Internet Protocol Version 4 (IPv4), has now been exhausted by its address space. Hence, a successor has been in increasing deployment, Internet Protocol Version 6 (IPv6). For reasons of actuality, IPv6 is considered in this paper. The IP protocol frames the message into a datagram and ensures its consistent and correct handling. In the physical world, this datagram corresponds to a piece of freight and its packaging, a π-container. Table 6 shows the attributes of this datagram and transforms these into the RBPI.

The routing table contains information about the next appropriate hub, as provided by the BGP-Protocol. For dynamic routing, this table has to be actualized with transport bandwidth information, represented by the free capacity of vehicles on the road that leads to a particular challenge.

*Table 6: Transformation of IPv6 datagram from DI to RBPI*

| IP attribute | # bit | Function within DI | Function within RBPI |
|---|---|---|---|
| Version | 4 | Version of the IP-Protocol, distinguished between IPv4 and IPv6. | Distinguishable versions of the *handling of π-containers* [disposable or reusable]. |
| Traffic Class | 8 | This field defines different priority levels in terms of quality of service. | Traffic class in the RBPI might signify requirements, like *special treatment* [temperature] and/or *express deliveries* [time]. |
| Flow Label | 20 | Contains the randomly selected identification number of a virtual end-to-end connection. | *Label packages that require extraordinary treatment*, for example the transport of animals. |
| Payload Length | 16 | Specifies how many bytes follow the header as so-called payload. | Specifies how much space or weight is required in the π-container, like *payload, available transport volume* or *max. number of pallets*. |
| Next Header | 8 | Mode indication 'connectionless' or 'connection-oriented'. | Specification of routing method. |
| Hop Limit | 8 | This attribute specifies the maximum number of network nodes through which a packet may pass before it is dropped. | This attribute can be transferred 1:1 to physical logistics, this would correspond to the *maximum number of intermediate π-nodes within the transport chain*. |
| Source/ Destination | 32/32 | Determines the source and destination address. | |

**Transformation of Layer 2: From Data Link Layer (OSI) to Link Layer (OLI)**

The Data Link Layer ensures *'a reliable and functioning connection between both the end device and the transmission medium'* (Badach and Hoffmann, 2007). It *'takes the datagram from the network layer and passes it from the current node to the next node in the network'* (Liesa et al., 2020). In order to prevent transmission errors and data loss, this layer contains functions for error detection, error correction and data flow control. The physical addressing of data packets also takes place on this layer.

*Table 7: Link Layer Transformation Table*

| Task within DI | Principle | Task within RBPI |
|---|---|---|
| **Framing** | | |
| Divides bitstream messages into manageable data units (frames). | Specialization | As every manageable unit must be transportable individually. It corresponds to the smallest unit, e.g. a pallet. |
| **Physical addressing** | | |
| The link layer adds a header to a frame containing the address of sender and/or receiver. | Instantiation | Corresponds to the delivery address of the shipment. |
| **Access control** | | |
| If two or more devices are using one link, protocols are necessary to determine which device has control over the link and when. | Analogy | Access to a bandwidth medium means access to appropriate π-movers with free capacity. Hence, remote access to the π-mover's cargo space must be ensured. |
| **Flow control** | | |
| If data is sent faster than it can be received, a flow control mechanism is used to avoid overwhelming the receiver. | Specialization | If a π-node is at its capacity or handling limit, transport vehicles on their way to the π-node could be informed in order to avoid congestion in the π-node. |
| **Error correction** | | |
| When a damaged or lost message is detected, it is resent. | Analogy | Lost freight components cannot simply be sent again, but must be reordered. If damaged or lost freight components cannot be reproduced from the current π-node (e.g. 3D-Print), they must be reordered from the beginning. |

On the DI, data packages that are handled are checked for damage. In the case that a package is either damaged or lost, it is simply generated and sent again from the current node. This leads to a problem with the transfer of this functionality into the world of physical objects. In some cases this can be compensated by generic part design at the hub location (3D print).

**Transformation of Layer 1: The Physical Layer**

The task of the Physical Layer in the DI is to *'ensure the transport of unformatted digital information units'* (Badach, Hoffmann and Knauer, 1997). Hence, this layer focuses on the electrical and mechanical properties of the transmission media. For this purpose, a transmission channel is provided via which information is exchanged, such as electrical cables, fiber optics or plug connections. Within the RBPI, the physical layer handles the movement of the physical objects in road-based vehicles. As discussed in (Kaup and Demircioglu, 2017), the correspondence to transmission media are the vehicles on the road. Hence, the bandwidth counterpart of transport media is the free capacity of vehicles moving on roads.

*Table 8: Physical Layer Transformation Table*

| Task within DI | Method | Task within RBPI |
|---|---|---|
| **Transport Channel Provision** | | |
| Bandwidth is given by the media, like electric wiring or fiber optics. | Analogy | Bandwidth definition correlates to free capacity of vehicles on the road. |
| Medium provisioning. | Analogy for π-movers | Load securing, so that freight is safely stowed in the vehicle. |
| **Amplification** | | |
| In order to prevent a signal from becoming so weak over long distances that it can no longer be interpreted, it is amplified in the network. | Instantiation for π-movers | Refueling π-movers with gasoline, electricity, compressed or liquid hydrogen. |
| | Analogy for π-containers | Preserve cargo, for example, from breaking the cold chain. Maintenance of services for π-containers, e.g. the provision of temperature or power control. |

Bandwidth is given in the DI by the media itself, like electric wiring or fiber optics. This corresponds in a one-to-one transformation to smaller streets (analogy to electric wiring) or highways (analogy to fiber optics). With respect to the RBPI this analogy is not best suitable as the term of bandwidth, because there the bandwidth correlates to free capacities of vehicles on the road. Nevertheless, the comparison to road traffic is dared or aspired here. This means that π-movers driving on a road with available capacities correspond to a transmission medium in the DI. In order to determine the transport bandwidth, information on the utilization rate of the driving π-movers is required (Kaup and Demircioglu, 2017). A second task of the physical layer is the amplification of signals (Badach, Hoffmann and Knauer, 1997) in order to prevent a signal from becoming more weak over long distances. An instantiation would lead to the refueling/-charging of vehicles or other kinds of π-movers, like starship robots. However, some freight requires special treatment, such as refrigeration. Hence, the maintenance of containers with temperature and/or power supply is the second analogy to the amplification of signals on the DI.

**Artifact Framework Design for RBPI**

All functionalities, methods and attributes, transferred from the DI to the RBPI in Section 3 lead to the design of a framework artifact, as shown in Figure 5. The corresponding functionalities of the DI in the RBPI were added to the five layer model from Section 2 that leads to the following artifact:

| Layer | DI | RBPI |
|---|---|---|
| 5 | Overarching Routing | Dynamic Routing Tables |
| 4 | De-/Fragmentation | De-/Fragmentation |
|   | Flow & Error Control (Macro) | Flow & Error Control (Macro) |
| 3 | Routing | Routing |
|   | Logical Addressing | Logical Addressing |
| 2 | Physical Addressing | Local Hub Addressing |
|   | Framing | Framing into PI-Containers |
|   | Flow & Error Control (Micro) | 3D-Print of lost freight items |
|   | Access Control | PI-Mover Access Control |
| 1 | Amplification | Container Ecosystem Provision |
|   |   | Bandwidth Monitoring |
|   | Transport Channel Provision | Cargo Securing |

*Figure 5: Artifact Framework for RBPI as a result of the Transformation*

The rounded rectangles on the left show the functions in the DI and those on the right the corresponding functionalities in the RBPI. When a solid frame of a rectangle is shown in Figure 5 on the RBPI side, a closely coupled principle was performed, such as specialization or instantiation. In the case of a loosely coupled analogy, the frame is shown dotted. The components shown in blue represent identified vehicle requirements. The circle with the (A) indicates that the load is automatically secured against slipping in the cargo space. The circle in which the (B) is located represents the feedback of the freely available capacities of the vehicles to the dynamic routing tables. The freely available transport capacity feedbacks might be replicated to all hubs in the system. This feedback loop is represented by the blue arrow. The prerequisite for this is that the vehicles have knowledge of their load states or their free capacities. This knowledge can either be implemented via an in-vehicle tracking system or via a fleet management system. The circle with the letter (C) represents the optional power supply of containers that provide the goods with special treatment, such as cooling.

## 5. Conclusions and further work

In the future, the Physical Internet will gain importance in transportation and logistics, not least for road-based transport. This paper identifies protocols of the Data Internet that are relevant for transport and transforms their properties and methods to the RBPI, which answers $SQ_1$. The result of this work is a framework artifact, which represents transferred and adapted methods of the DI for the RBPI. This design artifact answers $SQ_2$. From this point, vehicle-relevant requirements are derived in order to answer $SQ_3$. In future, vehicles should have an automatic load securing system, as well as be able to provide remote access to the cargo space. In order to be able to use real-time information for routing freight within the RBPI, vehicles must have information about their loading status and communicate these to the RBPI's routing entity. It is supposed that future work might consider how π-containers can provide an ecosystem for including freight, e.g. a cooling mechanism. For this purpose, a power supply between vehicles and π-containers has to be designed. The automotive industry as well as logistics operators are recommended to take these requirements into account in order to make vehicles and π-containers fit to the RBPI.